\documentclass[aps,pre,twocolumn,epsfig,floats]{revtex4-2}

\usepackage{newtxtext}
\usepackage{newtxmath}
\usepackage{bm}
\usepackage{amsmath}
\usepackage{physics}
\usepackage{color}
\usepackage{graphicx}
\usepackage{float}
\usepackage{bbold}

\usepackage{float}
\usepackage[utf8x]{inputenc}

\DeclareSymbolFont{myletters}{OML}{ztmcm}{m}{it}
\DeclareMathSymbol{\uplambda}{\mathord}{myletters}{"15}

\DeclareSymbolFont{cmlargesymbols}{OMX}{cmex}{m}{n}
\let\sum\relax
\DeclareMathSymbol{\sum}{\mathop}{cmlargesymbols}{"50}
\DeclareSymbolFont{cmletters}{OML}{cmm}{m}{it}
\SetSymbolFont{cmletters}{bold}{OML}{cmm}{b}{it}
\DeclareSymbolFontAlphabet{\mathnormal}{cmletters}

\usepackage{hyperref}
\hypersetup{
  colorlinks=true,
  citecolor=blue,
  linkcolor=blue,
  urlcolor=blue}

\newcommand{\raisedchi}{\raisebox{\depth}{\(\upchi\)}}

\begin{document} 

\title{Exploring the properties of quantum scars in a toy model}
 
\author{Sudip Sinha and S. Sinha} 
\affiliation{Indian Institute of Science Education and Research-Kolkata, Mohanpur, Nadia-741246, India}
 
\date{\today}

\begin{abstract}
We introduce the concept of ergodicity and explore its deviation caused by quantum scars in an isolated quantum system, employing a pedagogical approach based on a toy model. Quantum scars, originally identified as traces of classically unstable orbits in certain wavefunctions of chaotic systems, have recently regained interest for their role in non$-$ergodic dynamics, as they retain memory of their initial states. We elucidate these features of quantum scars within the same framework of this toy model. The integrable part of the model consists of two large spins, with a classical counterpart, which we combine with a random matrix to induce ergodic behavior. Scarred states can be selectively generated from the integrable spin Hamiltonian by protecting them from the ergodic states using a projector method. Deformed projectors mimic the `quantum leakage' of scarred states, enabling tunable mixing with ergodic states and thereby controlling the degree of scarring. In this simple model, we investigate various properties of quantum scarring and shed light on different aspects of many$-$body quantum scars observed in more complex quantum systems. Notably, the underlying classicality can be revealed through the entanglement spectrum and the dynamics of `out--of--time--ordered correlators'.
\end{abstract}


\maketitle

\section{Introduction}
Ergodicity is a fundamental concept in statistical mechanics that justifies the emergence of the Gibbsian prescription of thermodynamic equilibrium from microscopic dynamics~\cite{Sinai,Cornfield,Halmos}.
The ergodic evolution conveys the idea of exploring all possible configurations (microstates) given sufficient time, ensuring the equivalence between the time and ensemble average of the macroscopic thermodynamic quantities as well as the loss of memory of the initial condition~\cite{Boltzmann}.
Typically, when the system approaches a thermodynamic equilibrium, the~average of the macroscopic quantities can be described by an appropriate ensemble, a~process commonly referred to as `thermalization'.
Classically, to~study the dynamical behavior of a system, one typically investigates its phase space trajectories.
Although the deterministic evolution of an isolated system strongly depends on the initial conditions, the~memory of which is eventually erased during ergodic dynamics, allowing the system to explore the available region of the phase space determined by the conserved quantities. 
This concept is intuitively related to chaos theory, according to which the trajectories exhibit exponential sensitivity on the initial conditions~\cite{Strogatz_book}, leading to `phase space mixing'~\cite{Zaslavsky_chaos_book}. Therefore, chaos is believed to be a key ingredient for ergodicity in classical~systems.

On the other hand, extending the idea of chaos and ergodicity in a quantum system is not straightforward due to the absence of phase space trajectories, as~a consequence of the Heisenberg uncertainty principle. At~the same time, understanding thermalization in isolated quantum systems has also been a long$-$standing puzzle~\cite{Polkovnikov-review}. In~a quantum system, the~statistical properties of the eigenvalues as well as the eigenvectors play a crucial role in diagnosing the underlying chaos and ergodicity, utilizing the tools from the random matrix theory (RMT) \cite{Mehta-RMT,Haake2010}. Initially, the RMT was employed to capture the universal features of spectral statistics in complex quantum systems, such as atomic nuclei~\cite{Wigner_original1,Wigner_original2,Wigner_original3}. Later, the~Bohigas--Giannoni--Schmit (BGS) conjecture laid the basic foundation of quantum chaos by identifying a connection between the classically chaotic systems and spectral statistics of their quantum counterparts~\cite{BGS}.
According to this conjecture, the~spectral statistics of the Hamiltonian of a chaotic system follow the properties of the Gaussian orthogonal ensemble (GOE) of random matrices. Meanwhile, the~eigenstate thermalization hypothesis (ETH) has been proposed to explain thermalization, which is related to the statistical properties of individual eigenstates~\cite{Deutsch1991, Srednicki1994, Srednicki1999, Rigol2008, Polkovnikov-review, Santos-review, ETH_review_Deutsch2018}. Within~ETH, the~microcanonical thermalization can be explained for typical states with random components similar to the eigenvectors of a random matrix. From~this point of view, RMT plays a pivotal role in elucidating the hidden link between the underlying chaos, ergodicity, and thermalization, which is universal and not system$-$specific~\cite{Polkovnikov-review, Santos-review}.

Interestingly, not all complex systems are guaranteed to be ergodic. The~Fermi--Pasta--Ulam--Tsingou problem is one such example where, despite its complexity, the~system seems to evade ergodicity~\cite{FPU, Izrailev-rev-FPU, Tsingou}.
In recent years, identification of the factors that hinder thermalization and cause ergodicity breaking, has become an intense area of research~\cite{D_Huse_2007_MBL,MBL_quasiperiodic2013,Huse-rev-MBL,
Abanin2019,Schreiber2015,MBL_expt2016,Kinoshita2006,spin_glass_review1,
spin_glass_review2,Langer_spin_glass_review,quantum_glass_book}. One such factor, namely the many$-$body quantum scar (MBQS), has gained significant interest following an experiment on ultracold Rydberg atoms~\cite{Rydberg_expt_scar}, where certain specific initial configurations (N\'{e}el$-$like density wave state) demonstrated coherent revivals and slow relaxation to equilibrium, in~stark contrast to ergodic dynamics. This is remarkable in the sense that MBQS violates the Markovian property of ergodicity, which may have potential applications in quantum technologies~\cite{Pappalardi_metrology2023}.
Such revival phenomena has been attributed to the presence of a vanishingly small fraction of non$-$thermal states~\cite{Abanin_Rydberg_scar2018,Papic_Rydberg_scar2018,
Motrunich_Exact_Scar_Rydberg2019,Abanin_scars_review2021}, which remain protected from the surrounding ergodic states as a consequence of an emergent symmetry~\cite{Choi2019,spin1_XY_scars2019}. 
Furthermore, such symmetries can be associated with a spectrum generating algebra (SGA) \cite{SGA_book,Moessner_scars_review2022,Moudgalya_review2022}. 
The MBQS have already been extensively studied in a wide variety of interacting quantum systems, which not only include lattice models~\cite{Lukin2019,Abanin2020PRX,magnonscars2019,spin1_XY_scarsMPS2020,
AKLT_scar_Moudgalya2,AKLT_scar_Motrunich2020,AKLT_Shiraishi2019,
eta_pairing_Motrunich2020,correlated_hopping_BHM2020,
scars_optical_lattice2020,tilted_1d_FHM2021,Richter2023, Onsager_scars2020,lattice_gauge_scars2021, transverse_ising_ladder2020,Kitaev_chain2022,JJ_arrays2022,
Heisenberg_clusters2023} but also collective systems like Dicke model~\cite{ubiquitous_scarring}, its variants~\cite{BJJ_scar,OTOC_KDM,two_component_BJJ}, and~many more~\cite{Sinha_review,Evrard2023,Evrard2024,Pizzi2024,Lerose2024, Mueller2024,Omiya2024,CT_scars,CT_scars2,KCT_scars}.  
Other than Rydberg atoms, MBQS have also been experimentally realized in a superconducting processor~\cite{scars_scqubits2022} and a Bose$-$Hubbard quantum simulator~\cite{BHM_simulatorscar2022}.

Intuitively, the~single$-$particle wavefunctions of a non$-$interacting quantum chaotic system, such as the Bunimovich stadium, are expected to exhibit uniform spatial distribution with random fluctuations.
However, a~few higher energy eigenstates exhibit localization around the unstable classical orbits, which have been dubbed as `quantum scars'--reminiscent of the underlying unstable dynamics~\cite{Heller1984}.  In~the case of a many$-$body quantum system, an~analogous scenario of MBQS remains unclear due to the absence of well$-$defined phase space trajectories. Therefore, establishing a unified framework for non$-$ergodic behavior of MBQS and its connection with the underlying classicality is a challenging task, which deserves further investigation.
In this context, certain time dependent variational states such as the matrix product states (MPS) have recently been used to unfold an effective phase space dynamics~\cite{Lukin2019,Abanin2020PRX}. Interestingly, some MPS {\it Ans\"{a}tze} reveal isolated unstable periodic orbits in this space, corresponding to the long lived revival phenomena observed in the Rydberg chain~\cite{Lukin2019}. 
Furthermore, within~this effective phase space, MBQS maybe described by the regions of regular dynamics surrounded by a chaotic sea, which corresponds to `mixed phase space' behavior~\cite{Abanin2020PRX}.

The above$-$mentioned mixed phase space description can provide a naive visualization of the emergence of athermal states corresponding to MBQS.
From the viewpoint of Einstein--Brillouin--Keller (EBK) quantization~\cite{Wimberger_book}, the~regular regions (periodic orbits) of the mixed phase space can be thought of as the tower of protected states surrounded by the ergodic sea. 
In certain many$-$body systems, like the spin$-$1 XY model, the~tower of states remains perfectly protected~\cite{spin1_XY_scars2019}, which is associated with the spectrum generating algebra (SGA). When the initial configuration is a linear combination of these particular states, the~dynamics remain constrained to the protected subspace, leading to perfect revivals. 
However, in~more realistic (complex) systems like the PXP chain that describes the Rydberg atoms, the~protected states start mixing with the surrounding ergodic states as the associated algebra is deformed~\cite{Choi2019,Moessner_scars_review2022}, resulting in a `quantum leakage'. On~the other hand, due to the quantum leakage, the~exact dynamics is not restricted within the effective phase space defined by the time$-$dependent variational {\it Ans\"{a}tze} and eventually leave that subspace~\cite{Lukin2019,Abanin2020PRX}.
Consequently, there is a decay in the revival dynamics and the system relaxes, albeit much more slowly compared to ergodic dynamics. This suggests that quantum leakage is a key factor in controlling the degree of~scarring.

In this paper, we present a simple toy model that combines collective large spins with a random matrix Hamiltonian to generate the scarred states, which elucidates the different aspects of the scarring phenomena as well as unfolds its connection with the underlying classicality.
We consider the bare Hamiltonian to be integrable,  with~a well$-$defined classical counterpart and use a random matrix to trigger ergodic behavior. The~`projector method'~\cite{AKLT_Shiraishi2019,scar_Mori2017} is employed to embed exact eigenstates of the integrable part of the Hamiltonian within a sea of ergodic states, generating a tower of protected states.
We model quantum leakage by deforming the projection operators, causing the protected states to mix with the surrounding ergodic states. This method not only facilitates the study of scarring but also provides a way to control its degree by tuning the quantum leakage appropriately. 
In the present work, we explore the manifestation of quantum leakage in various properties of scars such as entanglement, revival, and~out--of--time--ordered correlator (OTOC) dynamics~\cite{Larkin,Shenker2014,Maldacena2016,Swingle-QFT-OTOC,Galitski2017,Sachdev2017,Swingle_unscrambling2018,
GarciaMata-OTOC, OTOC_classicalquantumDicke,Swingle_OTOC_tutorial,Swingle2016,
Lev2017,information_scrambling2020,OTOC_scars,
Richter_saturation,Ray-OTOC-MBL,Arul-scrambling,OTOC_anisotropicDicke,markovic_saturation,Fradkin-OTOC}. Importantly, we demonstrate how the underlying classicality can be unveiled from the entanglement spectrum (ES) and OTOC dynamics, which can be extended to more generic quantum~systems.

The rest of the paper is organized as follows: We describe the model of two large spins and discuss the properties of the protected subspace in Section~\ref{model_toy_ham}. In~Section~\ref{ergodicity_and_entanglement}, we discuss the ergodic properties of the systems and investigate the entanglement properties of the protected states. In~Section~\ref{quantum_dynamics}, we study the quantum dynamics of the system, with~a particular focus on the OTOCs, and~analyze the effect of the scarred states. Finally, we conclude with a discussion in Section~\ref{discussion}.

\section{Model}
\label{model_toy_ham}
To study the deviation from ergodicity due to the scarred states, we consider a toy model comprising two components, described by the following Hamiltonian: 
\begin{eqnarray}
\hat{\mathcal{H}} &=& \hat{\mathcal{H}}_{0} + \hat{\mathcal{H}}_{\rm I}, \label{main_Ham} \\
\hat{\mathcal{H}}_{0} &=& J\,(\hat{S}_{1z} + \hat{S}_{2z}), \label{bare_Ham} \\
\hat{\mathcal{H}}_{\rm I} &=& \hat{\mathcal{P}}(\raisedchi)\,\,\hat{\mathcal{H}}_{\rm GOE}\,\,\hat{\mathcal{P}}(\raisedchi),
\label{random_Ham}
\end{eqnarray}
where the integrable part of the Hamiltonian, $\hat{\mathcal{H}}_{0}$, describes the precession of two non$-$interacting large spins along the $z$$-$axis with frequency $J/\hbar$ and the operators $\hat{S}_{1z}=\hat{S}_{z} \otimes \mathbb{1}$, $\hat{S}_{2z} = \mathbb{1} \otimes \hat{S}_{z}$ represent the $z$$-$component of the corresponding spins with same magnitude $S$. Both the spin operators as well as the identity matrix ($\mathbb{1}$) have the dimension, $\mathcal{D}=2S+1$. 
Unless otherwise specified, we set $\hbar=1$ for the remainder of the paper.
We consider the eigenstates $\ket{m_{1z},m_{2z}} = \ket{m_{1z}} \otimes \ket{m_{2z}}$ of the Hamiltonian $\hat{\mathcal{H}}_{0}$ as the orthonormal basis states with dimension $\mathcal{N}=\mathcal{D}^2$, where $\ket{m_{1z}}$ and $\ket{m_{2z}}$ correspond to eigenstates of $\hat{S}_{z}$, satisfying the following eigenvalue equations:
\begin{eqnarray}
\hat{S}_{1z}\ket{m_{1z},m_{2z}} = m_{1z}\ket{m_{1z},m_{2z}},\\ 
\hat{S}_{2z}\ket{m_{1z},m_{2z}} = m_{2z}\ket{m_{1z},m_{2z}}.
\end{eqnarray}
In the second part, $\hat{\mathcal{H}}_{\rm I}$, induces ergodic behavior due to inclusion of a symmetric random matrix ($\hat{\mathcal{H}}_{\rm GOE}$) from Gaussian orthogonal ensemble (GOE). The~elements of these matrices are randomly sampled from the standard Gaussian distribution with zero mean and unit variance~\cite{Haake2010}. Notably, the~components of the eigenvectors of such matrices also follow a Gaussian distribution, which is a characteristic of typical ergodic states, in~agreement with ETH~\cite{Haake2010,Berry_conjecture}. Moreover, the~distribution of the spacing between the consecutive eigenvalues of the GOE matrices follows the Wigner surmise, reflecting their chaotic nature, in~accordance with the BGS conjecture. The~operator $\hat{\mathcal{P}}(\raisedchi)$ denotes a deformed projector that projects out the states $\ket{m_{1z},0}$, which can be expressed as follows:
\begin{eqnarray}
\hat{\mathcal{P}}(\raisedchi) &=& \mathbb{1} \otimes \mathbb{1} -\!\! \sum^{S}_{m_{1z}=-S}\!\!\!\raisedchi \ket{m_{1z},0} \bra{m_{1z},0},
\end{eqnarray}
where the strength of the deformation is given by $\raisedchi \in [0,1]$. 
While superposing the random part of the Hamiltonian $\hat{\mathcal{H}}_{\rm I}$ destroys the integrable nature of the spin Hamiltonian $\hat{\mathcal{H}}_{0}$, the~main purpose of the projector is to only protect the eigenstates $\ket{m_{1z},0}$ corresponding to the first spin from the ergodic states, effectively mimicking the scarring phenomena in an overall chaotic system. Moreover, the~underlying classicality of such scarred states can be explored due to the semiclassical nature of the large spins. For~$\raisedchi=1$, the~states $\ket{m_{1z},0}$ are perfectly projected out, whereas decreasing its value from unity leads to mixing with the surrounding ergodic states, giving rise to a quantum leakage. 
In a realistic quantum system, the~scarred states are not perfectly protected from the ergodic states due to such quantum leakage, which in the present model can be controlled by tuning the redefined parameter $\epsilon = 1-\raisedchi$.

As a consequence of an emergent symmetry, in~a typical interacting many$-$body system such as the spin$-$1 XY model, the~tower of protected scarred states is associated with a spectrum generating algebra (SGA) \cite{spin1_XY_scars2019,Moessner_scars_review2022,Moudgalya_review2022}.
For a Hamiltonian $\hat{\mathcal{H}}$, the~SGA is defined by the following relation:
\begin{eqnarray}
[\hat{\mathcal{H}},\hat{\mathcal{O}}^\dagger]\ket{\psi_{n}} = \omega\hat{\mathcal{O}}^\dagger \ket{\psi_{n}}, \label{RSGA}
\end{eqnarray}
where $\mathcal{O}^\dagger$ acts as an excitation operator that generates a tower of energy eigenstates $\ket{\psi_{n}}$ with fixed spacing $\omega$. From~the above relation in Equation~\eqref{RSGA}, it is evident that for a state $\ket{\psi_{n}}$ with eigenvalue $E_{n}$, $\mathcal{O}^\dagger$ generates the next eigenstate $\hat{O}^\dagger\ket{\psi_{n}}$ with energy eigenvalue $E_{n}+ \omega$, so that the tower of states is created by the consecutive action of the excitation operator on the lowest energy eigenstate $\ket{\psi_{0}}$. Note that, the~excitation operator is analogous to the creation operator $\hat{a}^\dagger$ in an one dimension harmonic oscillator, which follows the same SGA for all the energy eigenstates. In~the present model, the~operators corresponding to different spin components $\hat{S}_{1\alpha}$ ($\alpha=x,y,z$) satisfy the SU(2) algebra,
\begin{eqnarray}
[\hat{S}_{1\alpha},\hat{S}_{1\beta}] = \dot{\iota}\epsilon_{\alpha \beta \gamma} \hat{S}_{1\gamma}.
\end{eqnarray}
For the integrable part of the Hamiltonian $\hat{\mathcal{H}}_{0}$, the~ladder operator $\hat{S}_{1+} = \hat{S}_{1x}+\dot{\iota}\hat{S}_{1y}$ acts as the excitation operator $\hat{\mathcal{O}}^\dagger$, satisfying the SGA,
\begin{eqnarray}
[\hat{\mathcal{H}}_{0},\hat{S}_{1+}] = [J(\hat{S}_{1z} + \hat{S}_{2z}),\hat{S}_{1+}] = J[\hat{S}_{1z},\hat{S}_{1+}] = J\hat{S}_{1+}. \label{present_model_RSGA}
\end{eqnarray}
The successive operation of the operator $\hat{S}_{1+}$ on the eigenstate $\ket{\psi_{0}} =\ket{-S,0}$ generates a tower of equispaced states $\ket{\psi_{n}}=\ket{m_{1z},0}$ with energy spacing $J$. 
The inclusion of the random part of the Hamiltonian $\hat{\mathcal{H}}_{\rm I}$ deforms the SGA in Equation~\eqref{present_model_RSGA}, introducing corrections corresponding to the tower of states $\ket{\psi_{n}}$ as follows:
\begin{eqnarray}
[\hat{\mathcal{H}}_{\rm I},\hat{S}_{1+}]\ket{\psi_{n}} &=& \epsilon\,\hat{\mathcal{H}}_{\rm GOE}\ket{\psi_{n+1}} \notag\\
&-& \epsilon(1-\epsilon)\sum^{2S}_{i=0}\ket{\psi_{i}}\bra{\psi_{i}}\,\hat{\mathcal{H}}_{\rm GOE}\,\ket{\psi_{n+1}} \notag\\
&-& \epsilon\,\hat{S}_{1+}\hat{\mathcal{H}}_{\rm GOE}\ket{\psi_{n}}\notag\\
&+& \epsilon(1-\epsilon)\sum^{2S-1}_{i=0}\ket{\psi_{i+1}}\bra{\psi_{i}}\hat{\mathcal{H}}_{\rm GOE}\ket{\psi_{n}}.
\end{eqnarray}
Interestingly, the~correction term vanishes only for the perfect projectors with $\epsilon=0$, completely protecting the tower of states from the ergodic part of the spectrum. On~the other hand, the~presence of a quantum leakage for $\epsilon \neq 0$ deforms the algebraic structure of the tower of states, resulting in a deviation from their integrable nature. 
Following the above prescription using the deformed projector, a~tower of quasi$-$integrable states can be formed within the surrounding ergodic spectrum, which act as scarred states, leading to deviation from ergodicity. Note that, using the projector method, only a fraction of $1/(2S+1)$ scarred states are generated within the Hilbert space of dimension $\mathcal{N} = (2S+1)^2$, which becomes vanishingly small for $S\gg 1$. Therefore, the~overall ergodicity remains unaffected even in the presence of the scarred states.
Next, we explore the properties of the scarred states and their dynamical~manifestation.

\section{Entanglement and~Ergodicity}
\label{ergodicity_and_entanglement}
\begin{figure}
\centering
\includegraphics[width=\columnwidth]{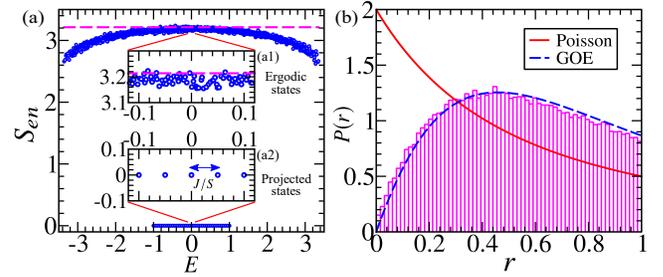}
\caption{Ergodic properties for $\epsilon = 0$ are as follows: (\textbf{a}) Variation in Entanglement entropy (EE), $S_{en}$, of~the energy eigenstates with energy density, $E \equiv \mathcal{E}_{n}/S$. The~pink dashed line corresponds to the page value of EE [see Equation~\eqref{page_formula}]. The~insets (\textbf{a1},\textbf{a2}) display the ergodic and projected states zoomed in a small window around $E$$\sim$0, respectively. (\textbf{b}) Distribution of the consecutive level spacing ratio $P(r)$ for 100 different realizations (ensembles) of random matrix $\hat{\mathcal{H}}_{\rm GOE}$, which agrees well with the GOE class as $\langle r \rangle$$\sim$0.523. We set $J=1$, $S=20$ in this and the rest of the figures unless otherwise~specified.}
\label{Fig1}
\end{figure}

In bipartite systems, entanglement is a quantum phenomenon that refers to the correlation between two subsystems of a larger system, such that the state of one subsystem cannot be described independently from the other. These correlations do not have any classical counterpart. 
In the present model, the~integrable part of the Hamiltonian $\hat{\mathcal{H}}_{0}$ comprises two non$-$interacting large spins which do not have any entanglement; therefore, the wavefunction corresponding to the entire system can be written as a product state describing the individual spins. However, the~inclusion of the random part $\hat{\mathcal{H}}_{\rm I}$ generates entanglement between the two spins, which can be quantified using the `entanglement entropy' (EE).
To compute the EE corresponding to the first spin sector which is our system of interest, we obtain its reduced density matrix $\hat{\rho}_{1}$ by tracing out the second spin sector from the total density matrix $\hat{\rho}=\ket{\psi}\bra{\psi}$ of some state $\ket{\psi}$ describing the composite system,
\begin{eqnarray}
\hat{\rho}_{1} = {\rm Tr}_{2}\hat{\rho}=\sum^{S}_{m_{2z}=-S}(\mathbb{1}_{1} \otimes \bra{m_{2z}})\,\hat{\rho}\,(\mathbb{1}_{1} \otimes \ket{m_{2z}}),
\end{eqnarray}
where we remind that $\{\ket{m_{2z}}\}$ is a set of orthonormal basis states of the second spin sector and $\mathbb{1}_{1}$ is the identity matrix corresponding to the first spin sector.
Following von Neumann's prescription, the~EE corresponding to the first spin sector is given by,
\begin{eqnarray}
S_{en} = -{\rm Tr}\hat{\rho}_{1}{\rm ln}{\hat{\rho}_{1}}. 
\end{eqnarray}
In thermodynamics, entropy is a measure of disorder in a system that quantifies the number of possible accessible microscopic configurations (microstates). 
At the classical level, intuitively, the~phase space mixing triggered by chaos can be reflected in the growth of entropy.
Analogously, the~EE can be viewed as a probe of randomness in a quantum system, thus serving as a useful metric for ergodicity~\cite{Neill2016,Rigol_EE,Srednicki_EE,Lewenstein-QKT}. 
In general, for~a random state of dimension $\mathcal{N}$, the~maximum limit of EE corresponding to a subsystem of size $\mathcal{N}_{A}$ is given by the page value~\cite{Page_value_EE},
\begin{eqnarray}
S_{page} = {\rm ln}\mathcal{N}_{\rm A}-\mathcal{N}^2_{A}/2\mathcal{N}. \label{general_formula_page}
\end{eqnarray}
One of the characteristic features of the scarred states is that they exhibit low EE compared to a typical ergodic state, reflecting their non$-$ergodic nature.
In the present model, for~$\epsilon=0$, the~projected states $\ket{m_{1z},0}$ are exact eigenstates of the Hamiltonian in Equation~\eqref{main_Ham}. 
Since these projected states can be written as a direct product $\ket{m_{1z},0} = \ket{m_{1z}} \otimes \ket{0}$, their reduced density matrices corresponding to the first spin sector $\hat{\rho}_{1} = \ket{m_{1z}}\bra{m_{1z}}$ represent pure states, and~the associated EE is zero, as~shown in Figure~\ref{Fig1}a.
The remaining eigenstates contribute to the ergodic part of the spectrum and approach the Page value of EE,
\begin{eqnarray}
S_{page} = {\rm ln}(2S+1)-1/2, \label{page_formula}
\end{eqnarray}
where $\mathcal{N}_{A}=2S+1$ and $\mathcal{N}=(2S+1)^2$. 
Apart from the bipartite EE, the~quantum Fisher information (QFI) \cite{Helstrom_book,Braunstein1994} has become another important tool for studying the signature of multipartite entanglement in quantum systems consisting of many interacting degrees of freedom~\cite{Smerzi2009,Smerzi2012,Toth2012}, such as the spin chains, which can have potential applications to quantum metrology~\cite{Braunstein1994,Lloyd2011,Petz_book,Smerzi2018, Pappalardi_metrology2023}. 
In particular, the~QFI has also been used for capturing the signature of scarring phenomena in the PXP chain, which indicates a long$-$range order in the scarred states~\cite{QFI_scar}.

As previously mentioned, since the number of such projected states only contribute a small fraction of the total dimension of the Hilbert space, $2S+1 \ll \mathcal{N}=(2S+1)^2$ for $S \gg 1$, the~overall ergodicity does not deviate significantly and spectral analysis of the total Hamiltonian follows that of a GOE class of random matrix, reflecting its overall chaotic nature~\cite{Haake2010}.
One of the methods of analyzing the spectral statistics involves distribution of the consecutive level spacing ratios~\cite{D_Huse_2007_MBL,Bogomolny_2013},
\begin{eqnarray}
r_{i} = \frac{{\rm min}(s_{i},s_{i+1})}{{\rm max}(s_{i},s_{i+1})},
\end{eqnarray}
where $s_{i}=\mathcal{E}_{i}-\mathcal{E}_{i+1}$ denote the nearest$-$neighbor level spacings.
For regular dynamics, the~distribution is given by,
\begin{eqnarray}
 P(r)=\frac{2}{(1+r)^2},
\end{eqnarray}
and the corresponding average value $\langle r \rangle$$\sim$0.386 \cite{D_Huse_2007_MBL,Bogomolny_2013}. Whereas in the chaotic regime, the~distribution is given by,
\begin{eqnarray}
P(r) = \frac{27}{4}\frac{(r+r^2)}{(1+r+r^2)^{5/2}},
\end{eqnarray}
with the corresponding average value $\langle r \rangle$$\sim$0.53 \cite{D_Huse_2007_MBL,Bogomolny_2013}.
In the present case, as~illustrated in Figure~\ref{Fig1}b, we observe that the distribution $P(r)$ agrees well with GOE statistics and $\langle r \rangle$$\sim$0.523, indicating that the overall chaotic behavior of the Hamiltonian $\hat{\mathcal{H}}$ remains unaffected even in the presence of a small fraction of athermal~states.

\begin{figure*}
\centering
\includegraphics[width=\textwidth]{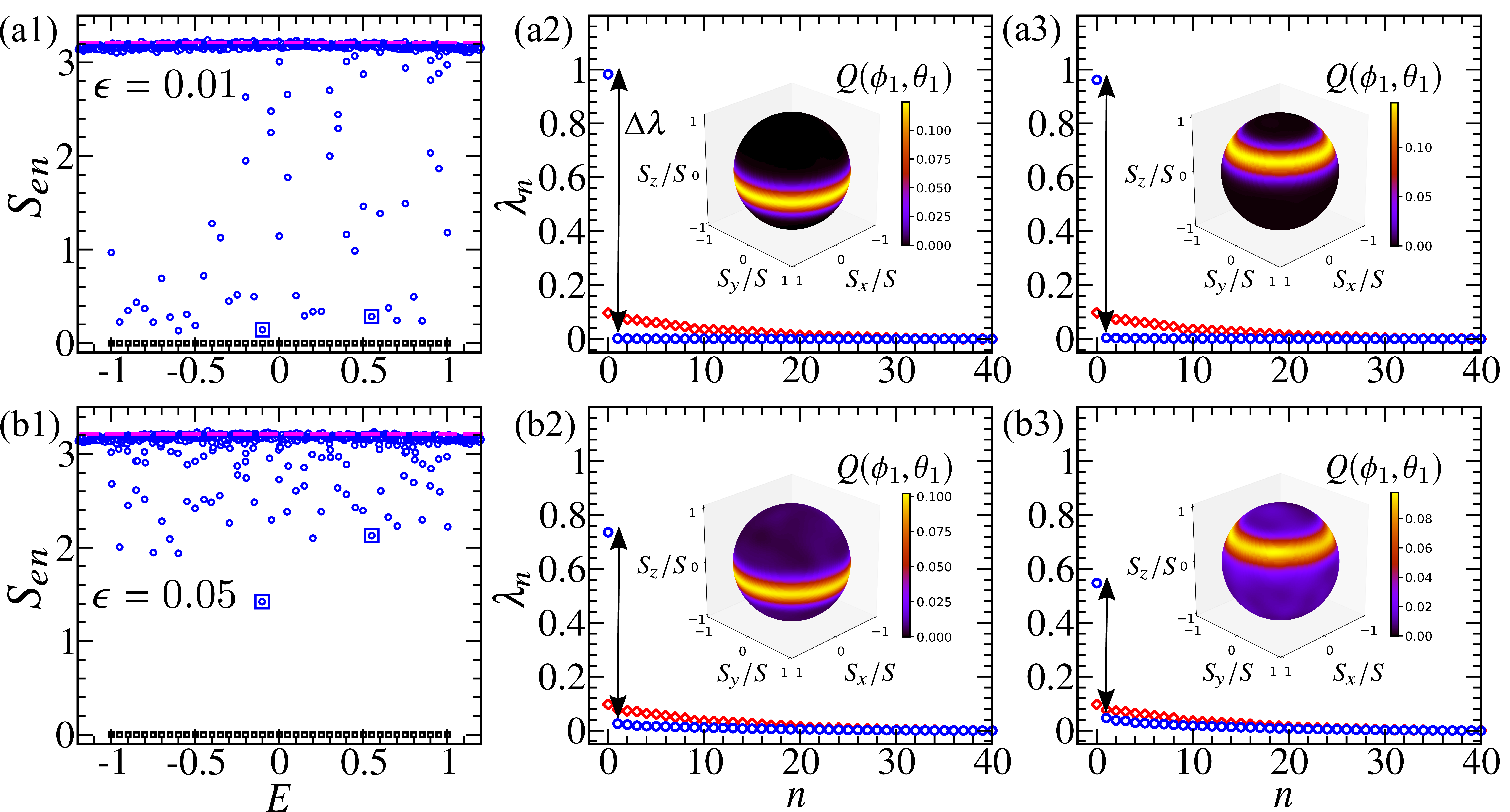}
\caption{The properties of the scarred states for a single realization of random matrix: (\textbf{a1},\textbf{b1}) EE of the energy eigenstates for different values of $\epsilon$. The~black squares denote EE of the eigenstates for $\epsilon = 0$. Entanglement spectrum (ES) and Husimi distribution corresponding to the first spin sector of the energy eigenstates (blue circles) marked in (\textbf{a1},\textbf{b1}) for (\textbf{a2},\textbf{a3}) $\epsilon$$\sim$0.01 and (\textbf{b2},\textbf{b3}) $\epsilon$$\sim$0.05, respectively. The~double headed arrows mark the gap, $\Delta\uplambda$, separating the largest eigenvalue and the extended tail in the ES. The~red diamonds correspond to the ES of a random eigenstate of a GOE matrix of the same dimension, reflecting the ergodic~behavior.}
\label{Fig2}
\end{figure*}

As previously mentioned, for~perfect projectors $\epsilon = 0$, the~reduced density matrices of the projected states represent pure states corresponding to the eigenstates $\ket{m_{1z}}$ of the integrable Hamiltonian, $\hat{\mathcal{H}}_{1} = J\hat{S}_{1z}$, describing the first spin.
It is evident from the above equation that $\hat{S}_{1z}$ serves as a conserved quantity since it commutes with $\hat{\mathcal{H}}_{1}$, and~the Heisenberg equations of motion for the other spin components are given by,
\begin{eqnarray}
\dot{\hat{S}}_{1\pm} = \pm\dot{\iota}J \hat{S}_{1\pm}. \label{Heisenberg_eq}
\end{eqnarray}
In the limit $S\gg 1$, the~spin components can be treated classically describing a large spin~vector,
\begin{eqnarray}
\vec{S} = S(\sin{\theta}\cos{\phi},\sin{\theta}\sin{\phi},\cos{\theta}), \label{classical_spin_vect}
\end{eqnarray}
where $(\phi,\theta)$ denote its orientation on the Bloch sphere, corresponding to the angular coordinates, with~$\theta$ being the polar angle from the $z$$-$axis and $\phi$ being the azimuthal angle from the $x$$-$axis in the $x$$-$$y$ plane, respectively. 
From Equation~\eqref{Heisenberg_eq}, the~corresponding semiclassical equations of motion of the spin vector is given by,
\begin{eqnarray}
\dot{\phi} = J, \quad  \frac{d}{dt}\cos{\theta} = 0,
\end{eqnarray}
where $\phi$ and $\cos{\theta}$ are canonical conjugate variables.
It follows from the above equation that the~motion of the classical spin vector corresponds to the periodic orbits, $\phi(t)=Jt$ with fixed $\theta$, which is analogous to spin precession around the direction of a constant applied magnetic field. The~different eigenstates $\ket{m_{1z}}$ of $\hat{\mathcal{H}}_{1}$ are related to these classical periodic orbits with $\cos{\theta}=m_{1z}/S$. Such underlying classicality can be captured using the spin coherent states~\cite{Radcliffe1971}, which are appropriate semiclassical representation for such classical spin vector [Equation~\eqref{classical_spin_vect}], maintaining the minimum uncertainty. 
The spin coherent state corresponding to orientation $(\phi,\theta)$ can be written as follows:
\begin{eqnarray}
\ket{\phi,\theta} = \cos^{2S}\!{(\theta/2)}\exp\left(\tan(\theta/2)e^{\dot{\iota}\phi}\hat{S}_{-} \right)\ket{S},
\end{eqnarray}
where $\ket{S}$ is the eigenvector of $\hat{S}_{z}$ with largest eigenvalue $S$.
To visualize the classical structure of such spin states, we utilize the Husimi distribution $Q(\phi,\theta)$, describing the semiclassical phase density. The~Husimi distribution corresponding to the reduced density matrix $\hat{\rho}_{1}$ is defined as follows:
\begin{eqnarray}
Q(\phi,\theta) = \bra{\phi,\theta}\hat{\rho}_{1}\ket{\phi,\theta},
\end{eqnarray}
that reduces to $|\bra{\phi,\theta}\ket{m_{z}}|^2$ for the perfectly projected states $\ket{m_{1z},0}$, involving the overlap with the coherent states $\ket{\phi,\theta}$, yielding the probability distribution over the Bloch sphere.
As evident from Figure~\ref{Fig2}(a2,a3), the~phase space densities are indeed localized around the classical orbits, revealing the underlying classical integrable nature of the projected scarred states. 
In the presence of deformation with $\epsilon \ll 1$, the~quantum leakage allows mixing with the ergodic states. Consequently, the~projected states are no longer the exact eigenstates of $\hat{\mathcal{H}}_{1}$, nevertheless, they still retain the underlying classical structure, which is reflected in their Husimi distribution [see Figure~\ref{Fig2}(b2,b3)]. However, with~increasing $\epsilon$, the~localized phase space density begins to spread, and~eventually, the~localized structure is washed out for sufficiently large $\epsilon$, indicating the ergodic~nature.

\begin{figure}
\centering
\includegraphics[width=\columnwidth]{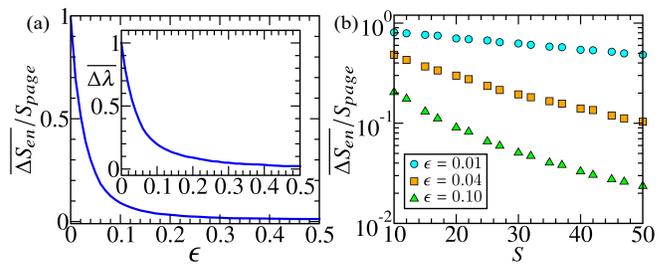}
\caption{Degree of scarring: (\textbf{a}) Variation in the average deviation of the EE $\overline{\Delta S}_{en}$ scaled by the maximum limit (page value), and~the average gap $\overline{\Delta \uplambda}$ in the ES (inset) at energy density $E$$\sim$0 with increasing deformation strength $\epsilon$, for~$S=20$. (\textbf{b}) Variation in $\overline{\Delta S_{en}}/S_{page}$ with increasing large spin magnitude $S$ for different values of $\epsilon$. The~averaging (indicated by $\bar{\boldsymbol{\cdot}}$ ) is performed in two steps: first, over~the states with minimum EE in a small energy density window $\Delta E = 0.09/S < 1/S$ around $E$$\sim$0, and~then over an ensemble of (\textbf{a}) 100 and (\textbf{b}) 1000 random matrices $\hat{\mathcal{H}}_{\rm GOE}$.}
    \label{Fig3}
\end{figure}

Although the classical correspondence of the integrable part of the Hamiltonian $\hat{\mathcal{H}}_{1}$ enables a direct visualization of the underlying classical nature of the projected scarred states through the Husimi distribution, a~deeper insight can also be gained by closely examining their entanglement properties.
We analyze the entanglement spectrum (ES) i.e., the set of eigenvalues $\{\uplambda_{m}\}$ of the reduced density matrix,
\begin{eqnarray}
\hat{\rho}_{1} = \sum^{2S}_{n=0} \uplambda_{n} \ket{n}_{1}\bra{n}_{1},
\end{eqnarray}
where $\ket{n}_{1}$ are the eigenvectors of $\hat{\rho}_{1}$. This can also be equivalently obtained from the Schmidt decomposition of a state $\ket{\psi}$ \cite{Ekert_AJP1995},
\begin{eqnarray}
\ket{\psi} = \sum_{n} \sqrt{\uplambda_{n}}\ket{n}_{1} \otimes \ket{n'}_{2},
\end{eqnarray}
where $\{\ket{n}_{1}\}$ ($\{\ket{n'}_{2}\}$) form an orthonormal basis state of the first (second) sector and $\uplambda_{n}\geq 0$ determine the degree of entanglement between the two spins which satisfy $\sum_{n}\uplambda_{n}=1$.
For the perfectly projected states, their reduced density matrix $\hat{\rho}_{1}=\ket{m_{1z}}\bra{m_{1z}}$ represents a pure state having only one non$-$vanishing unit eigenvalue corresponding to the eigenvector $\ket{m_{1z}}$. A~qualitative change in the ES can be noticed by tuning the deformation parameter $\epsilon \ll 1$, introducing the quantum leakage. Apart from the few large eigenvalues (closer to unity), an~extended tail in the spectrum is generated, and~both the components are separated by a spectral gap $\Delta \uplambda$, as~shown in Figure~\ref{Fig2}(a2,a3,b2,b3). Even in the presence of mixing with ergodic states, the~eigenvectors corresponding to the few large eigenvalues in the ES retain the information about the integrable states and the Husimi distribution of the density matrix constructed from them exhibits the corresponding classical structure~\cite{two_component_BJJ}. On~the other hand, the~extended tail of the ES reflects the ergodic component that arises from mixing. With~increasing the quantum leakage by tuning $\epsilon$, the~spectral gap $\Delta \uplambda$ decreases and the spectrum is eventually filled with the continuously  distributed eigenvalues, similar to that of an ergodic state (like the typical eigenvector of $\hat{\mathcal{H}}_{\rm GOE}$). 
Therefore, the~spectral gap in the ES carries the signature of the underlying classicality, separating a few large eigenvalues from the extended tail, retaining the information about the integrable part of the Hamiltonian.
This spectral gap has also been identified for quantum scars in realistic interacting systems, such as the two$-$component Bose--Josephson junction (BJJ) \cite{two_component_BJJ} and Heisenberg spin clusters~\cite{Heisenberg_clusters2023}.
From the structure of the ES, it is evident that the EE of the projected scarred states increases [see Figure~\ref{Fig2}(a1,b1)] as the spectral gap $\Delta \uplambda$ decreases due to increasing quantum leakage. 
In the presence of quantum leakage, the~non$-$ergodic behavior associated with the degree of scarring can be quantified using both the deviation of the EE from the maximum value (page limit), $\Delta S_{en}=S_{page}-S_{en}$, as~well as the spectral gap $\Delta \uplambda$. Additionally, the~spectral gap in the ES can also serve as a link between non$-$ergodicity and the underlying classicality. 
We consider the states at the center of the energy band at energy density $E\sim0$, which exhibit minimum EE for the projected states and compute their average deviation scaled by the maximum limit (page value), $\overline{\Delta S}_{en}/S_{page}$, as~well as their average spectral gap $\overline{\Delta \uplambda}$, with~increasing deformation strength $\epsilon$. Here, the~averaging (indicated by $\bar{\boldsymbol{\cdot}}$) is performed in two steps: first, over~the different states within the energy density window, and~then over a large ensemble of random matrices $\hat{\mathcal{H}}_{\rm GOE}$ with same distribution. 
It is evident from Figure~\ref{Fig3}(a), both the quantities decay rapidly with $\epsilon$, indicating the onset of ergodic behavior for stronger deformation strengths. 
Similar phenomena have also been observed for quantum scars in BJJs~\cite{BJJ_scar,two_component_BJJ} and the coupled top model~\cite{CT_scars2}, where the degree of scarring reduces as the system approaches the ergodic regime by tuning the relevant~parameters.

We also analyze how the degree of scarring, due to the mixing phenomena, depends on the Hilbert space dimension by increasing the large spin magnitude $S$. For~this purpose, we compute the variation in the scaled quantity $\overline{\Delta S_{en}}/S_{page}$ for the projected scarred states with $S$ for different values of the deformation parameter $\epsilon$. As~shown in Figure~\ref{Fig3}(b), such deviation decreases with increasing values of $S$. Moreover, the~reduction in $\overline{\Delta S_{en}}/S_{page}$ is enhanced with increasing $\epsilon$, which indicates the influence of the dimensionality of the Hilbert space on the sensitivity of the deformation~parameter.

As seen from this example, the~entanglement properties of the scarred states elucidate the link between the non$-$ergodic features and the underlying classicality of the integrable component of the Hamiltonian. Such non$-$ergodic features and the underlying classicality can also be explored from the non$-$equilibrium dynamics, which we discuss in the next~section. 

\begin{figure*}
\centering
\includegraphics[width=\textwidth]{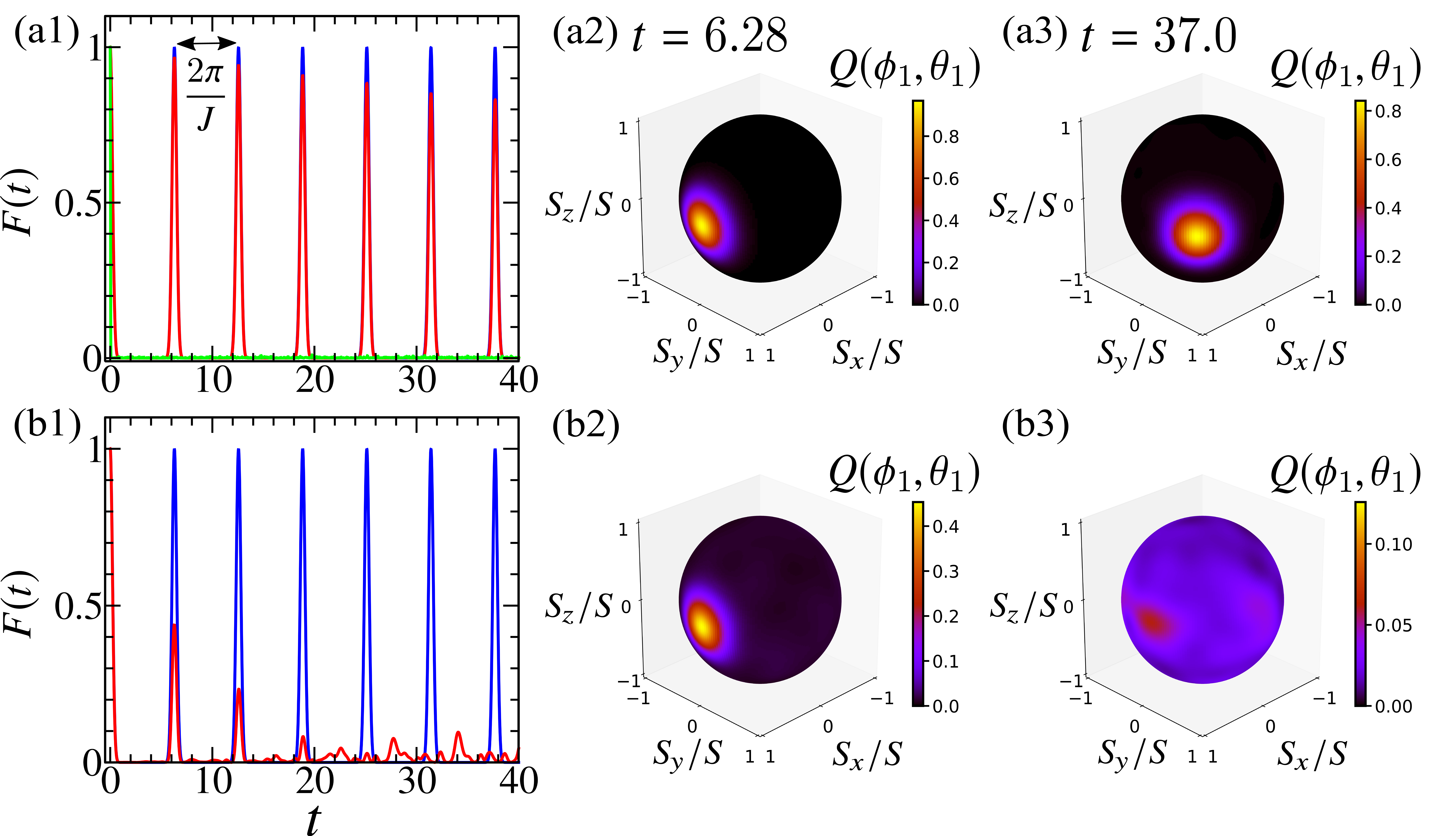}
\caption{Quantum dynamics starting from the initial state $\ket{\phi=0,\theta=\pi/2} \otimes \ket{m_{2z}=0}$ at $E$$\sim$0 : Time evolution of survival probability $F(t) =|\langle \psi(t) | \psi(0) \rangle|^2$ for (\textbf{a1}) $\epsilon = 0.01$ and (\textbf{b1}) $\epsilon = 0.05$  (red lines), compared with the dynamics for $\epsilon = 0.0$ (blue line). Snapshots of Husimi distribution corresponding to the first spin sector at different times for (\textbf{a2},\textbf{a3}) $\epsilon = 0.01$ and (\textbf{b2},\textbf{b3}) $\epsilon = 0.05$. The~green line in (\textbf{a}) exhibits a rapid relaxation of $F(t)$ starting from the initial state $\ket{\phi=0,\theta=\pi/2} \otimes \ket{m_{2z}=1}$ for $\epsilon=0$, indicating a complete loss of memory of the initial state (ergodic dynamics), as~it is not a linear combination of the perfectly projected~states.}
\label{Fig4}
\end{figure*}

\section{Signature of Quantum Scarring from~Dynamics}
\label{quantum_dynamics}
The main feature of quantum scars is associated with the non$-$ergodic dynamics, retaining the memory of the initial state, resulting in a revival phenomena. A~convenient way to study such behavior is to compute the survival probability,
\begin{eqnarray}
F(t) = |\bra{\psi(t)}\ket{\psi(0)}|^2,
\end{eqnarray}
which quantifies the overlap with the initial state $\ket{\psi(0)}$. Here, we consider the initial state as a product state having non$-$vanishing overlap with the projected scarred states,
\begin{eqnarray}
\ket{\psi(0)} = \ket{\phi=0,\theta=\pi/2} \otimes \ket{m_{2z}=0} \label{initial_state}
\end{eqnarray}
where $\ket{\phi=0,\theta=\pi/2}$ is a coherent state corresponding to the first spin representing a point $(\phi=0,\theta=\pi/2)$ of the periodic orbit at the equator of the Bloch sphere. As~evident from Figure~\ref{Fig4}(a1), for~$\epsilon = 0$, we observe perfect periodic revivals of $F(t)$ with a time period $\tau = 2\pi/J$ corresponding to the classical orbits, indicating that the dynamics is only constrained within the protected subspace. In~the presence of quantum leakage with $\epsilon \ll 1$, such revivals still persist; however, the amplitude of the oscillations decay with time [see Figure~\ref{Fig4}(b1)]. For~sufficiently strong deformation, the~state eventually becomes ergodic, resulting in a rapid decay of $F(t)$ without any revivals, indicating a complete loss of memory of the initial state. This behavior is further illustrated from the snapshots of the Husimi distributions at different times in Figure~\ref{Fig4}(a2,a3,b2,b3), showing the dynamics and spreading of the localized phase space wavepacket over the Bloch~sphere.

\begin{figure}
\centering
\includegraphics[width=\columnwidth]{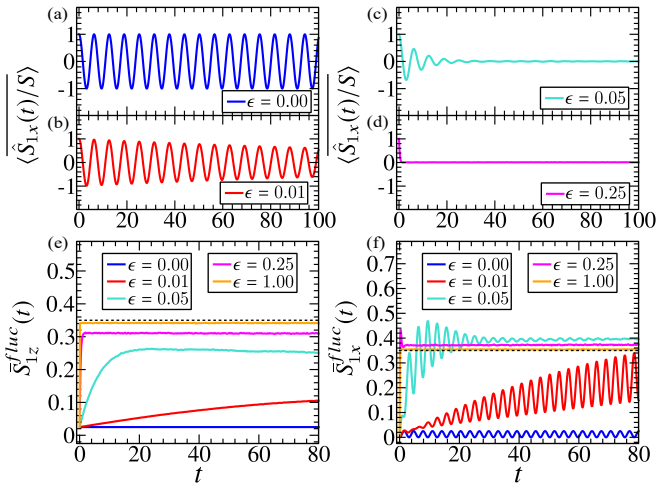}
\caption{Quantum dynamics starting from the initial state $\ket{\phi=0,\theta=\pi/2} \otimes \ket{m_{2z}=0}$ at $E$$\sim$0: (\textbf{a}--\textbf{d})~Time evolution of $\langle \overline{\hat{S}_{1x}(t)/S} \rangle$ for different values of $\epsilon$. Fluctuation dynamics of the spin operators (\textbf{e})~$\bar{S}^{fluc}_{1z}(t)$ and (\textbf{f}) $\bar{S}^{fluc}_{1x}(t)$. The~dotted line denotes the microcanonical value given by Equation~\eqref{spin_op_mc}. Note that, $\bar{\langle \boldsymbol{\cdot} \rangle}$ and $\bar{\boldsymbol{\cdot}}$ represent averaging over an ensemble of 100 random~matrices.}
\label{Fig5}
\end{figure}

To supplement the above analysis, we also study the time evolution of the different components of the first spin $\hat{S}_{1\alpha}/S$ ($\alpha = x,y,z$) as well as their fluctuations. Starting from the same initial state given in Equation~\eqref{initial_state}, we compute $\langle \overline{S_{1\alpha}/S} \rangle$, where $\bar{\langle \boldsymbol{\cdot} \rangle}$ represents the averaging over an ensemble of random matrices $\hat{\mathcal{H}}_{\rm GOE}$. As~depicted in Figure~\ref{Fig5}(a), for~$\epsilon = 0$, $\langle \overline{\hat{S}_{1x}(t)/S} \rangle$ exhibits perfect coherent oscillations with time period $\tau$, whereas $\langle \overline{\hat{S}_{1z}(t)/S} \rangle$ remains constant (which vanishes for an initial coherent state on the equator) since the operator $\hat{S}_{1z}$ is a conserved quantity. For~$\epsilon \ll 1$, the~coherent oscillations in $\langle \overline{\hat{S}_{1x}(t)/S}\rangle$ decay while the time period remains the same, as~shown in Figure~\ref{Fig5}(b,c). For~larger values of $\epsilon$, the~oscillations in $\langle \overline{\hat{S}_{1x}(t)/S} \rangle$ eventually cease to exist exhibiting a rapid decay [see Figure~\ref{Fig5}(d)].  To~capture the spreading of the phase space density, we also study the fluctuation dynamics of the spin operators,
\begin{eqnarray}
S^{fluc}_{1\alpha}(t) = \langle \hat{S}^{2}_{1\alpha}(t)/S^2 \rangle - \langle \hat{S}_{1\alpha}(t)/S\rangle^2,
\end{eqnarray}
where an averaging is performed over a large ensemble of random matrices.
As shown in Figure~\ref{Fig5}(e,f), in~the absence of deformation, $\bar{S}^{fluc}_{1z}(t)$ remains constant and does not grow since $\hat{S}_{1z}$ is a conserved quantity. On~the contrary, the~deformation in the projector leads to a growth in  $\bar{S}^{fluc}_{1\alpha}(t)$, the~rate of which increases with the increasing values of $\epsilon$, indicating mixing with the ergodic states.
Moreover, the~fluctuation dynamics is closely linked to the dynamics of fidelity OTOC (FOTOC), which has been used for detecting the underlying dynamical instability in collective quantum systems~\cite{AM_Rey_FOTOC,Lea_FOTOC,Corney_FOTOC,KCT_scars}, as~well as diagnosis of scarred states~\cite{KCT_scars}.
For sufficiently strong mixing, it is observed that $\bar{S}^{fluc}_{1\alpha}(t)$ saturates to a value equivalent to the microcanonical ensemble,
\begin{eqnarray}
\bar{S}^{fluc}_{1\alpha}(t\rightarrow \infty) &\sim& \langle \hat{S}^2_{1\alpha}/S^2 \rangle_{mc} - \langle \hat{S}_{1\alpha}/S \rangle^2_{mc}\notag\\ 
&\sim& \frac{1}{3}\left(1+\frac{1}{S}\right), \label{spin_op_mc}
\end{eqnarray}
reflecting the ergodic nature. Therefore, the~saturation value obtained from the fluctuation dynamics can also serve as a measure of degree of~ergodicity.

An alternative approach for diagnosing chaos in a generic quantum system employs the dynamics of out--of--time--ordered correlators (OTOC) \cite{Larkin,Shenker2014,Maldacena2016,Swingle-QFT-OTOC,Galitski2017,
Sachdev2017,Swingle_unscrambling2018,GarciaMata-OTOC,
OTOC_classicalquantumDicke,Swingle_OTOC_tutorial,Swingle2016,
Lev2017,information_scrambling2020,OTOC_scars,Richter_saturation,
Ray-OTOC-MBL,Arul-scrambling,OTOC_anisotropicDicke,
markovic_saturation,Fradkin-OTOC}, which has recently become popular for its application in black hole thermalization~\cite{Shenker2014,Maldacena2016} and quantum information theory~\cite{Swingle2016,Lev2017,information_scrambling2020,OTOC_scars}.
In a classical system exhibiting chaotic dynamics, it is well known that the separation between trajectories starting from initial conditions, subjected to a small perturbation $\delta q(0)$, grows exponentially with time, $\delta q(t) = e^{\Lambda t}\delta q(0)$, where $\Lambda$ is the Lyapunov exponent (LE), quantifying the degree of chaos.
This relation can also be derived from the Poisson Bracket of the position ($q$) and momentum ($p$) variables, as~follows:
\begin{eqnarray}
\{q(t),p(0)\}_{\rm PB} = \partial q(t)/\partial q(0) = e^{\Lambda t},
\end{eqnarray}
where the Poisson Bracket is defined as $\{A,B\}_{\rm PB}=\partial A/\partial q(0)\ \partial B/\partial p(0) - \partial A/\partial p(0)\ \partial B/\partial q(0)$. Its quantum analogue can be obtained by using the canonical quantization relation,
\begin{eqnarray}
\{q(t),p(0)\} \Leftrightarrow -\dot{\iota}\hbar[\hat{q}(t),\hat{p}(0)],
\end{eqnarray}
where $[\,\,,\,]$ denotes the commutator of the corresponding operators. Using this classical correspondence, one can expect to diagnose chaos from the following real quantity,
\begin{eqnarray}
C(t) = \langle [\hat{q}(t),\hat{p}(0)]^\dagger [\hat{q}(t),\hat{p}(0)] \rangle, \label{squared_commutator}
\end{eqnarray}
where $\langle \boldsymbol{\cdot} \rangle \equiv {\rm Tr}(\hat{\rho}_{0}\boldsymbol{\cdot})$, with~$\hat{\rho}_{0}$ being the density matrix corresponding to a suitably chosen initial state. 
In the semiclassical limits, the~underlying chaos can be quantified from the growth of $C(t)$ \cite{Shenker2014,Maldacena2016,Swingle-QFT-OTOC,Galitski2017,Sachdev2017,Swingle_unscrambling2018,GarciaMata-OTOC,Swingle_OTOC_tutorial,
OTOC_classicalquantumDicke,Swingle2016,Lev2017,OTOC_KDM}. The~squared commutator in Equation~\eqref{squared_commutator} can be generalized as~follows:
\begin{eqnarray}
C(t) = \langle [\hat{W}(t),\hat{V}(0)]^\dagger [\hat{W}(t),\hat{V}(0)] \rangle, \label{otoc_squared_commutator}
\end{eqnarray}
which is commonly referred to as the~OTOC.

\begin{figure}
\centering
\includegraphics[width=0.85\columnwidth]{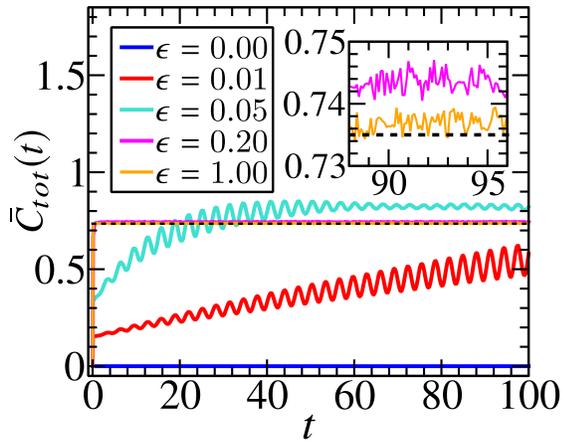}
\caption{OTOC dynamics: Time evolution of $\bar{C}_{tot}(t)$ for eigenstate with minimum entanglement in a small energy density window $\Delta E = 0.045 < 1/S$ around $E$$\sim$0 . The~inset shows the OTOC dynamics zoomed at long time and the dotted line denotes the microcanonical value given by Equation~\eqref{otoc_mc}. Note that, the~OTOC dynamics is averaged over 100 ensembles of random~matrices.}
\label{Fig6}
\end{figure}

Since the projected scarred states retain the memory of the integrable spin Hamiltonian, the~dynamics of OTOC of the spin operators can be utilized to distinguish them from the surrounding ergodic states with underlying chaos. Here, we consider the OTOC corresponding to the different components of the spin operators $\hat{S}_{1\alpha}$ and investigate the time evolution of their sum, which is defined as follows:
\begin{eqnarray}
C_{tot}(t) = -\!\!\sum_{\alpha = x,y,z} \langle \left[ \hat{S}_{1\alpha}(t),\hat{S}_{1\alpha}(0)\right]^2\rangle /S^4, \label{spin_microcanonical_otoc}
\end{eqnarray}
where the squared commutator is scaled by $S^4$ (equivalent to scaling the individual spin operators by $S$) to analyze its behavior in the semiclassical limit.
We evaluate the above$-$mentioned quantity for the projected scarred states and then perform an ensemble averaging over the random matrices ($\bar{C}_{tot}(t)$). As~expected, for~a completely projected state with $\epsilon = 0$, $\bar{C}_{tot}(t)$ remains vanishingly small and does not display any growth, confirming their non$-$chaotic nature as they retain the memory of the integrable part of the Hamiltonian [see Figure~\ref{Fig6}]. In~the presence of quantum leakage with $\epsilon \ll 1$, the~growth rate of $\bar{C}_{tot}(t)$ increases. Most importantly, it exhibits oscillation of period $\tau/2$, where $\tau$ is the time period of the underlying classical periodic orbits corresponding to the integrable spin Hamiltonian $\hat{\mathcal{H}}_{1}$. 
Such behavior has also been previously observed for scars of unstable periodic orbits in the coupled top model~\cite{CT_scars,CT_scars2}.
Even for many$-$body systems such as the PXP model, the~OTOC dynamics have been shown to exhibit periodic oscillations for the scarred states~\cite{OTOC_scars}.
In the ergodic regime with $\epsilon \rightarrow 1$, the~oscillations in $\bar{C}_{tot}(t)$ disappear completely and it attains saturation following a rapid growth, indicating the destruction of the scarred nature. In~a generic quantum system, the~OTOC of the finite dimensional operators typically attains a saturation value after a certain time. Apart from the growth rate of OTOC, its saturation value can also serve as a measure of degree of ergodicity for such systems~\cite{GarciaMata-OTOC,Richter_saturation,
Ray-OTOC-MBL,Arul-scrambling,OTOC_anisotropicDicke,markovic_saturation,
OTOC_KDM}. In~the ergodic regime, the~saturation value of the OTOC $\bar{C}^{sat}_{tot}$ can be estimated by considering the random nature of the eigenstates. For~the eigenstates of a random matrix Hamiltonian $\hat{\mathcal{H}}_{\rm GOE}$, the~reduced density matrix $\hat{\rho}_{0}$ resembles to that of a microcanonical ensemble, which is proportional to the identity matrix. In~this regime, the~OTOC approaches the steady state value, $C(t) \rightarrow \langle \hat{S}^2_{1\alpha}/S^2 \rangle^2$, as~the correlator ${\rm Tr}\big(\hat{S}_{1\alpha}(t)\hat{S}_{1\alpha}(0)\hat{S}_{1\alpha}(t)\hat{S}_{1\alpha}(0)\hat{\rho}_{0}\big)$ vanishes rapidly. For~typical ergodic states, $\langle \hat{S}^2_{1\alpha}\rangle$ converges to its statistical average in the microcanonical ensemble, $\langle \hat{S}^2_{1\alpha} \rangle_{mc} = S(S+1)/3$. As~a result, the~saturation value of the OTOC for the scaled variables can be expressed as follows:
\begin{eqnarray}
\bar{C}^{sat}_{tot} = \sum_{\alpha} \langle \hat{S}^2_{1\alpha}/S^2 \rangle^2_{mc} =  \frac{2}{3}\left(1+\frac{1}{S}\right)^2. \label{otoc_mc}
\end{eqnarray}
The deviation of the saturation value of OTOC from the above limit indicates a departure from ergodicity. Here, we emphasize that while the growth rate and saturation value of the OTOC can serve as measures of the degree of ergodicity, the~periodic oscillations observed in the OTOC during an initial time interval reveal the underlying periodic orbits associated with the scarred~states.
 
In addition, the~Krylov complexity method has recently emerged as an important tool for the detection of chaos in a generic interacting quantum system~\cite{K_op_complexity_PRX,K_state_complexity,K_complexity_JHEP1, K_complexity_JHEP2} which is related to the growth of an operator, and~also has connections to OTOC~\cite{K_op_complexity_PRX}. Moreover, it can describe the time evolution of an initial state within an optimal choice of basis, also referred to as `Krylov space'~\cite{K_state_complexity}. Notably, in~contrast to the growth of the Krylov complexity for typical ergodic states, it exhibits oscillatory behavior for the scarred states in the PXP chain~\cite{K_complexity_scar}, which is similar to the OTOC dynamics of scarred states observed in the present model. This indicates that the information of the scarred state mostly remains within the Krylov space, which can leak to the thermal part of the spectrum due to the mixing with the surrounding ergodic~states.

\section{Discussion}
\label{discussion}
To summarize, this work aims to extract the universal features of quantum scarring phenomena, independent of system$-$specific details, using a simple toy model based on two large spins. Additionally, it elucidates the connection between deviation from ergodicity and the underlying classicality of quantum scars. 
The Hamiltonian describing the model consists of two components: (i) Integrable part comprising of two large non$-$interacting spins with a well$-$defined semiclassical limit and (ii) random matrix component inducing the ergodic behavior. Ergodicity of an isolated quantum system is related to the statistical properties of the eigenstates, which resemble with eigenvectors of a random matrix, exhibiting maximum delocalization in any generic basis. On~the other hand, the~eigenstates of an integrable Hamiltonian exhibit localization of the semiclassical phase space density around the classical periodic orbits, which can be visualized from the corresponding Husimi distributions. However, such localized structure of the eigenstates is affected by appropriately mixing a random matrix to the integrable part of the Hamiltonian. By~utilizing the ‘projector method’, we selectively generate a tower of non$-$thermal scarred states that are exact eigenstates of the integrable part of the Hamiltonian. The~projectors are chosen in a manner such that only a small fraction of the eigenstates corresponding to one of the large spins, remain protected from the surrounding ergodic states. As~a consequence, the~overall chaotic nature of the system remains unaffected as the spectral statistics still follow the Wigner surmise. The~scarred states are identified from their entanglement entropy (EE) as they exhibit maximum deviation from the Page value compared to the ergodic states. We introduce a deformation in the projectors which causes a mixing between the scarred and the ergodic states in a controlled manner, mimicking the quantum leakage observed in more complex many$-$body systems~\cite{Lukin2019,Abanin2020PRX}. Notably, the~tower of scarred states follow a spectrum generating algebra (SGA) which becomes imperfect in the presence of such quantum leakage. Such phenomena of quantum leakage is manifested as spreading of the semiclassical phase space densities around the underlying periodic orbits, which in turn enhances the degree of ergodicity. This also gives rise to a decay in the coherent revival dynamics, which is the main feature of many$-$body quantum scars (MBQS). For~a generic scarred state, the~mixing between the integrable and the ergodic components can be effectively illustrated through the entanglement spectrum (ES) which exhibits a significant gap, separating a few large eigenvalues from the extended tail. Furthermore, the~integrable component of the state is encoded within the eigenvectors corresponding to larger eigenvalues, and~the effective density matrix constructed from them, exhibits localization of the semiclassical phase space density around the periodic orbits. On~the contrary, the~tail part resembles the ES of a typical ergodic state, which becomes more prominent with increasing quantum leakage, reflecting enhanced ergodicity. Thus, the~gap separating the two components in the ES can serve as an indicator of the degree of scarring, which can be controlled by a tunable quantum leakage. Such feature of the ES has  been first observed in a two$-$component Bose--Josephson junction (BJJ) \cite{two_component_BJJ}, as~well as in an interacting spin system~\cite{Heisenberg_clusters2023}.
The dynamical signature of the protected states is reflected from  the coherent revivals in the survival probability and oscillations in the time evolution of physical quantities such as the spin components. These revivals gradually decay due to mixing with the surrounding ergodic states.
This suggests that the memory of the initial state is still retained as the dynamics remain mostly confined to the subspace of the projected states, which is eventually lost due to the quantum leakage. Importantly, while the dynamics of the out--of--time--ordered correlator (OTOC) is widely used as a diagnostic tool for chaos in quantum systems, it can also be effectively employed to capture the signature of scars~\cite{OTOC_KDM,CT_scars,CT_scars2,OTOC_scars}. Specifically, it highlights two key aspects of the scarring phenomena: (i) slower growth rate and deviation of its saturation value from that of a typical ergodic state, indicating a departure from ergodicity; (ii) periodic oscillations for a certain duration of time before saturation, where the oscillating frequency corresponds to that of the classical periodic orbits, manifesting the underlying classicality. 
These features have also been observed for quantum scars of unstable periodic orbits in the coupled top model~\cite{CT_scars,CT_scars2}.

In the present work, this simple model not only elucidates the different aspects of MBQS but also sheds light on its connection with the underlying classical dynamics. Even if a semiclassical phase space description is not straightforward for an interacting many$-$body system, features like gap in the ES~\cite{two_component_BJJ,Heisenberg_clusters2023} and oscillations in the OTOC \mbox{dynamics~\cite{CT_scars,CT_scars2,OTOC_scars}} may still unfold the underlying classicality, which can serve as a useful tool for their detection. 
From this point of view, the~collective systems can provide an ideal platform to explore the classical route to scarring phenomena~\cite{Sinha_review} using the classical$-$quantum correspondence, which has been investigated in physical systems, such as the Dicke model~\cite{ubiquitous_scarring,OTOC_KDM}, BJJs~\cite{BJJ_scar,two_component_BJJ}, and~other large spin models~\cite{Evrard2023,Evrard2024,Pizzi2024,Lerose2024,Mueller2024, Omiya2024,CT_scars,CT_scars2, KCT_scars, Sinha_review}. In~general, the~scarred states are mainly identified from the deviation of the entanglement entropy in the ergodic regime and for systems with well$-$defined semiclassical limit, the~quantum scars of periodic orbits have also been identified~\cite{two_component_BJJ,Evrard2023,Evrard2024,Pizzi2024, Mueller2024, CT_scars,CT_scars2, KCT_scars,Omiya2024}, which is captured in this toy~model.

While we focused on quantum scars generated by uniform deformation of the projectors, this approach can be easily extended to generate scarred states in an energy$-$dependent manner, targeting specific regions of the eigenspectrum, particularly corresponding to the classical energies of the steady states of the integrable Hamiltonian. 
Furthermore, the~integrable part of the Hamiltonian can be treated as a collection of spin$-$1/2 systems, allowing the analysis of the behavior of multipartite entanglement of generic scarred states using quantum Fisher information.
Exploring the behavior of Krylov complexity for the detection of the scarred states is another direction of research, which has already been investigated in the PXP chain~\cite{K_complexity_scar}. It exhibits an oscillatory behavior that has also been observed in the OTOC dynamics of the scars discussed in the present model.
Moreover, the~present analysis reveals that the essential feature of the underlying classicality of the scarred states is contained within a set of few eigenvectors of the reduced density matrix corresponding to the larger eigenvalues. This suggests the applicability of the time$-$dependent variational states within a smaller subspace, such as the matrix product states with lower dimensionality, to~capture the scarring phenomena~\cite{Lukin2019,Abanin2020PRX}. It also indicates a close connection to the Krylov method which also generates a reduced space in the full Hilbert space, describing the time evolution of an appropriate initial state. 
From this point of view, the~quantum leakage can be though of as information exiting the Krylov space, which can also be probed from the growth of the Krylov complexity.
Additionally, it would also be interesting to study the localization of the scarred states as well as the growth of Shannon entropy in the Krylov basis, which can be important for their dynamical~identification.


\end{document}